# Superconductivity depression in ultrathin $YBa_2Cu_3O_{7-\delta}$ layers in $La_{0.7}Ca_{0.3}MnO_3$ / $YBa_2Cu_3O_{7-\delta}$ superlattices.

Z. Sefrioui [a)], M. Varela [a)], V. Peña [a)], D Arias [a) &)], J.E. Villegas [b)], W. Saldarriaga [c)], P. Prieto [c)], C. León [a)] J. L. Martínez [d)] and J Santamaría [a)]

[a)] GFMC, Departamento de Física Aplicada III, U Complutense de Madrid, 28040 Madrid, Spain
[b)] Departamento de Fisica de los Materiales, U. Complutense de Madrid, 28040 Madrid, Spain
[c)] Departamento de Fisica, Universidad del Valle A. A. 25360 Cali, Colombia
[d)] Instituto de Ciencia de Materiales de Madrid (ICMM-CSIC), Cantoblanco 28049 Madrid, Spain
[&] On leave from Universidad del Quindio, Armenia, Colombia.
Corresponding author: J. Santamaria, email: jacsan@fis.ucm.es.

**ABSTRACT**

We report on the depression of the superconducting critical temperature of ultra thin $YBa_2Cu_3O_7$ (YBCO) layers, when their thickness is reduced in presence of $La_{0.7}Ca_{0.3}MnO_3$ (LCMO) magnetic layers in [LCMO (15 u.c.)/ YBCO(N u.c.)] superlattices. The thickness of the manganite layer is kept at 15 unit cells and the YBCO thickness is varied between N=12 and N=1 unit cells. The structural analysis using x-ray diffraction and electron microscopy shows sharp interfaces with little structural disorder. While a critical temperature, Tc=85 K is found for 12 YBCO unit cells, superconductivity is completely suppressed for YBCO layer thickness below 3 unit cells. The possible interaction between superconductivity and magnetism is investigated.

Ferromagnetic (F) / superconductor (S) heterostructures have recently attracted much interest for applications in spin injection (three terminal) devices [1]. High Tc superconductors (HTS) and colossal magnetoresistance (CMR) oxides are interesting candidate materials because the low carrier density of the HTS and the almost full spin polarization of the CMR oxides can be combined to yield high sensitivity (gain) fast devices. A reduction of the critical current consistent with suppression of superconductivity by spin polarized quasiparticle injection has been reported by several groups in recent years [2,3], opening the door to practical devices based on complex oxides. The samples reported so far involve quite thick (50-100 nm ) YBCO layers, thus shadowing interface effects. However, interface properties are expected to play a dominant role in the physics of CMR/HTS F/S heterostructures, and extrinsic (interface alloying or roughness) or intrinsic factors (proximity effect) may deeply influence the performance of the devices. The use of superlattices (instead of bilayers) allows an in depth characterization of the interfaces with conventional structure probes like x-ray diffraction (XRD) or transmission electron microscopy (TEM). The presence of magnetism and superconductivity in this kind of samples has been reported before [4,5]. In this letter we report on the growth of LCMO/YBCO superlattices with ultrathin (1 to 12 unit cells) YBCO layers and fixed LCMO thickness (15 unit cells), to investigate how robust is the superconductivity of the YBCO when its thickness is reduced in presence of magnetic layers. We have found that superconductivity is depressed in presence of the adjacent LCMO layers. A structural analysis with XRD and TEM is used to explore the influence of interface disorder on the depression of the superconductivity.

Samples were grown in a high pressure (3.4 mbar) pure oxygen sputtering system at high temperatures (900 ºC). Individual YBCO films on STO (100) were fully epitaxial with Tc of 90 K and transition widths smaller than 0.5 K. Growth conditions, optimized for the YBCO, yielded LCMO single films with a ferromagnetic transition temperature $T_{CM}$= 200 K, and a saturation magnetization $M_S$= 400 emu/cm$^3$, close to the bulk value. Superlattices were grown keeping LCMO thickness fixed at 15 unit cells per period and systematically changing YBCO thickness from 1 to 12 unit cells. Samples were checked for the presence of magnetism and superconductivity by transport (resistivity) and susceptibility (SQUID) measurements.

Figure 1 shows resistance curves for a series of superlattices with increasing YBCO thickness. It can be observed that the superconductivity is completely suppressed for YBCO layer thickness of 1 and 2 unit cells. For larger YBCO layer thickness, however, Tc displays a monotonic increase up to a value of 85 K (close to that of thick single films) for N=12. Samples were magnetic with $T_{CM}$ close to 150 K and a saturation magnetization of 100 emu/ cm$^3$. Figure 2 displays a field cooled (FC) susceptibility measurement (to avoid Meissner effect shielding) in various fields, showing a clear magnetic transition. A hysteresis loop measured at 90 K (well above the superconducting transition) is shown as an inset in the same figure. All samples showed similar magnetic response.

Although the $T_c$ depression could indicate the interaction between superconductivity and magnetism, several extrinsic factors have to be discarded beforehand. In particular, a deficient oxygenation of the YBCO and interface disorder, roughness, strain or interdiffusion, could also cause the superconductivity depression. A deficient oxygenation of the YBCO through the manganite layers can be ruled out since the thickest (above 10 unit cells) YBCO layers almost completely recover the bulk critical temperature. A detailed structural characterization of the interface structure is necessary in order to exclude roughness or interdiffusion. The artificial modulation of the superlattices provides many satellite peaks in x-ray diffraction patterns, which are very sensitive (width and height) to interface disorder like steps or interdiffusion.

X-ray diffraction patterns of the same samples are shown in figure 3. Clear Bragg peaks (labeled (005) and (006)) can be observed in the figure. It is worthwhile to notice that since LCMO c lattice parameter is roughly 1/3 than that of YBCO, (003), (006), etc. peaks display enhanced superlattice contrast, whilst (004), (005) etc., are affected by finite size effects. In fact the width of the (005) peak is a good measure of the YBCO layer thickness. In addition clear satellites characteristic of the superlattice modulation can be observed. The inset of this figure shows a plot of the modulation length, Λ, obtained from satellite spacing versus the nominal YBCO thickness. The very good fit to a straight line demonstrates the accurate control of the deposition rate. X ray patterns were checked for the presence of interface disorder using the SUPREX 9.0 refinement software [6]. Refinements yielded a layer thickness fluctuation (step disorder roughness) of the manganite layer of 0.5-0.7 unit cells. Epitaxial mismatch strain is known to cause deep structural modifications, which have been proposed as a source of superconductivity depression in the cuprate superconductors [7]. However, we also did not found indications of epitaxial mismatch strain as expected from the small lattice mismatch between YBCO and LCMO: x ray refinement did not show changes in the intracell distances along the c direction.

We want now to take a close look at a sample with very thin (one unit cell) YBCO layer to explore the possibility of interdiffusion. If interdiffusion is occuring during high temperature growth one expects reduced superlattice contrast in these samples, contrary to what is observed (see figure 2). Figure 4 (a) shows a detailed view of the x ray diffraction pattern of a [LCMO (15 u. c.)/YBCO (1 u.c.)] together with the refinement. Many superlattice peaks can be observed around the superlattice Bragg peaks indicating very small interdiffusion (if any). The calculated spectrum, which is really close to the experimental data (line) was only containing step disorder (not interdiffusion) at the interface consisting in 0.5-0.7 manganite unit cells. In fact, the incorporation of small amounts (<10%) of La into Y sites considerably deteriorated the agreement of the calculated spectra. Although the x-ray refinement was consistent with the absence of interdiffusion, we can not completely rule out the substitution of Cu atoms by Mn, especially in the first YBCO perovskite blocks. Calculated spectra are expected to be rather insensitive to this interdiffusion due to the similar electron densities of Mn and Cu ($Z_{Cu}$=29, $Z_{Mn}$=25). It has been reported that the substitution of Cu by magnetic elements like Ni or Co into the chains reduces the carrier concentration (and accordingly the critical temperature) [8]. Figure 4(b) shows a TEM cross section view of the same superlattice with a single YBCO unit cell, obtained in a Philips CM200 microscope operated at 200 kV. Very flat interfaces can be observed, which given the small thickness of the layer is consistent with the absence of interdiffusion.

At this point it seems very unlikely that the systematic depression of the critical temperature when the YBCO thickness is reduced might result of extrinsic factors like deoxygenation or roughness. Another source of Tc depression in ultrathin layers is the reduced dimensionality [9]. In this context, we compare the depression of the critical temperature when the YBCO thickness is reduced in presence of magnetic (LCMO) and non magnetic ($PrBa_2Cu_3O_7$) layers with similar (fixed) thickness. Figure 5 shows that the critical temperature is further reduced in presence of LCMO magnetic spacers than of PBCO. This is an indication of the interaction between magnetism and superconductivity. Two different scenarios can be invoked to discuss this phenomenon: one is F/S proximity effect and the other is pair breaking by spin polarized carriers injected from the LCMO. In the F/S proximity effect, Cooper pairs diffusing into the magnetic layer are broken by the exchange interaction. A Tc depression results over relatively long length scales in the superconductor, which has been theoretically addressed by Radovic et al [10], and experimentally observed in metallic superlattices by several groups [11-13]. In addition since the current is injected through the upper YBCO layer, and although current distribution may be complex in these kind of samples, we can not exclude the possibility of pair breaking by polarized quasiparticles injected into the superconductor [1-3]. Further work will be necessary to explore the relative importance of both contributions.

In summary, we have presented the growth of high quality LCMO/YBCO superlattices with ultrathin YBCO layers. A structural analysis using XRD and TEM has shown sharp interfaces with little structural disorder. We have found a depression of the critical temperature when the YBCO thickness is reduced which strongly indicates the interaction between magnetism and superconductivity.

**Acknowledgments**

Work supported by CICYT MAT2000-1468, CAM 07N/0008/2001and Fundación Ramón Areces.

**Figure Captions:**

**Figure 1:** Resistance curves of a [LCMO$_{15u.c.}$/YBCO$_{N\ u.c.}$] superlattices with N= 1,2,3,4, 5, 6, 7, and 12 YBCO unit cells (from top to bottom).

**Figure 2.** FC magnetic moment vs temperature of [LCMO$_{15u.c.}$/YBCO$_{5\ u.c.}$] with applied fields of H=50, 100 and 500 Oe applied parallel to the layers. Inset: magnetization loop at T=90 K.

**Figure 3:** High Angle XRD pattern of [LCMO$_{15u.c.}$/YBCO$_{N\ u.c.}$] superlattices with N= 1,2,3,4, 5 and 7 YBCO unit cells (from top to bottom). Inset: modulation length as a function of nominal number of YBCO unit cells.

**Figure 4** (a) High Angle XRD pattern of [LCMO$_{15u.c.}$/YBCO$_{1\ u.c.}$] superlattice. The bottom curve is a SUPREX fit with 0.7 unit cells layer fluctuation in the LCMO and no interdiffusion. (b) TEM image of the same superlattice.

**Figure 5:** Critical temperatures for [LCMO$_{15u.c.}$/YBCO$_{N\ u.c.}$] (circles) and [PBCO$_{5u.c.}$/YBCO$_{N\ u.c.}$] (squares) superlattices. The thickness of the non superconducting spacer is the same in both cases.

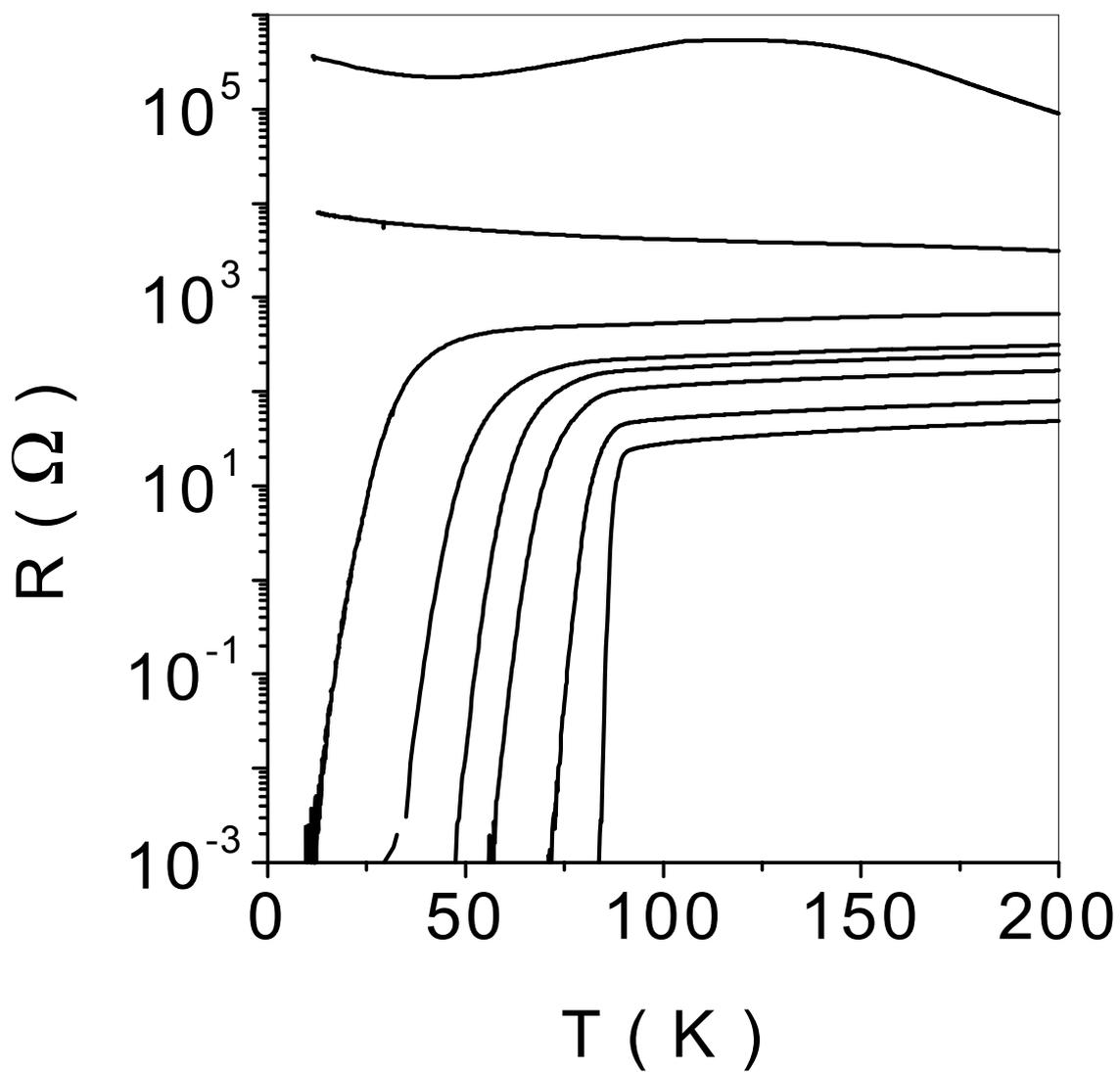

Z. Sefrioui figure 1

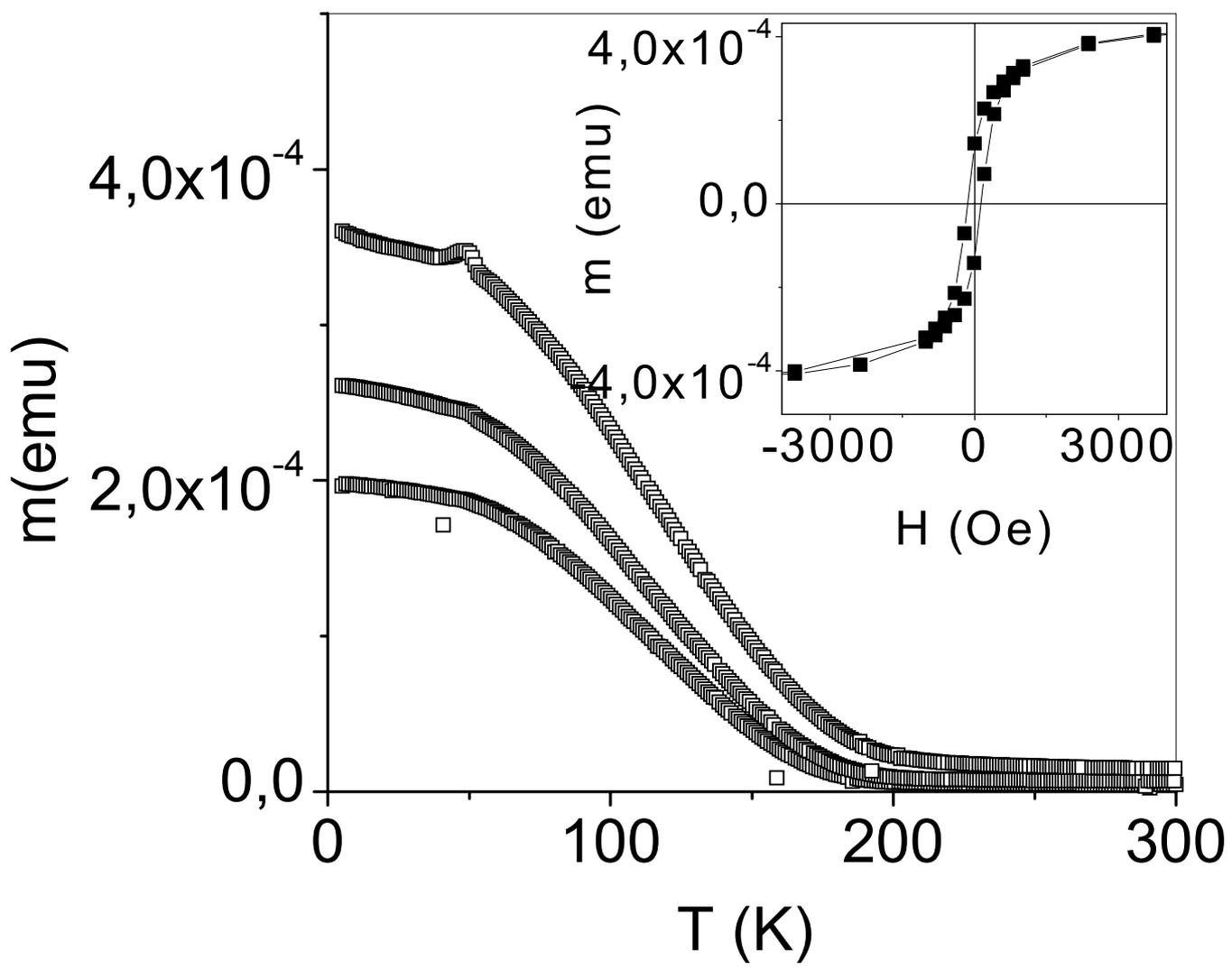

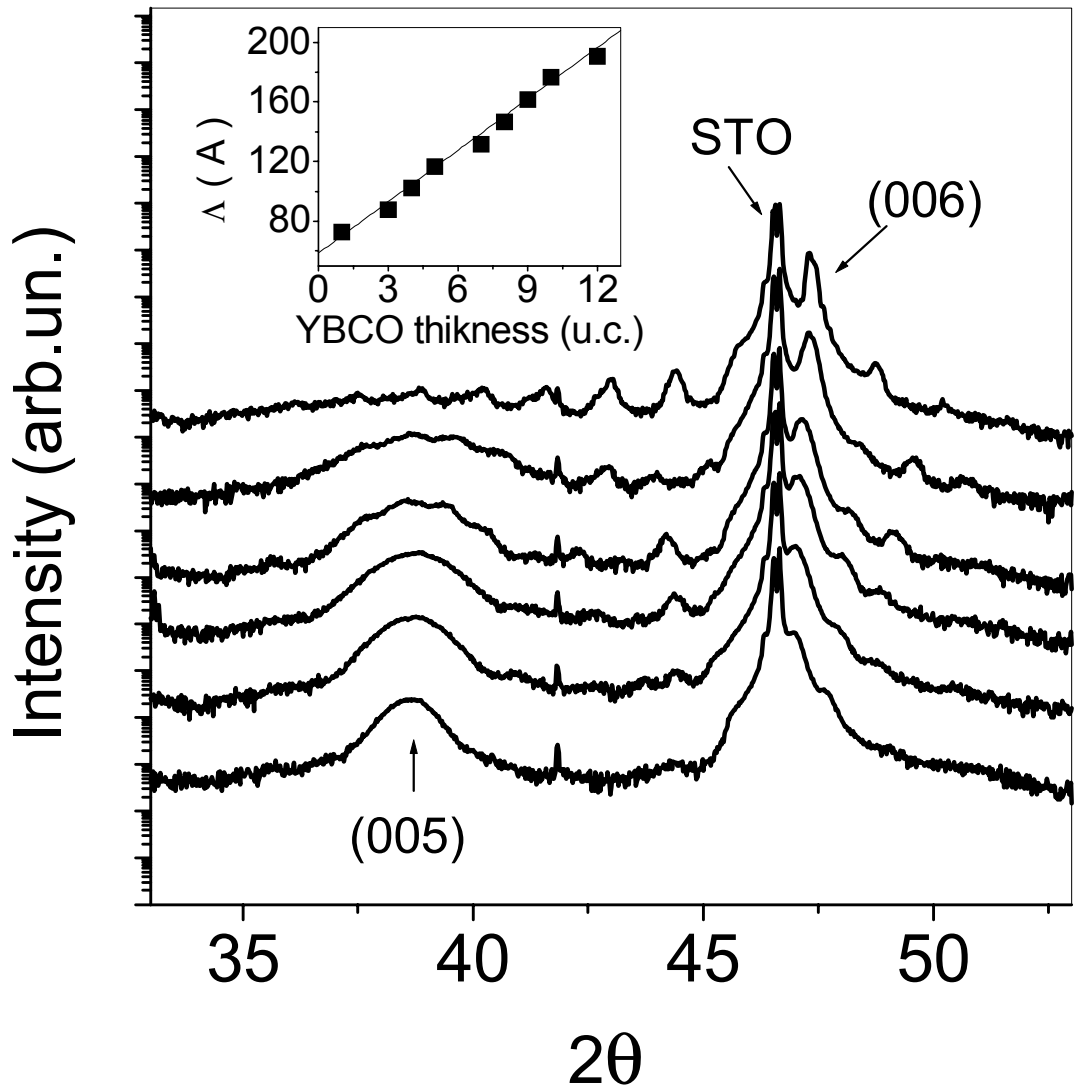

Z. Sefrioui figure 3

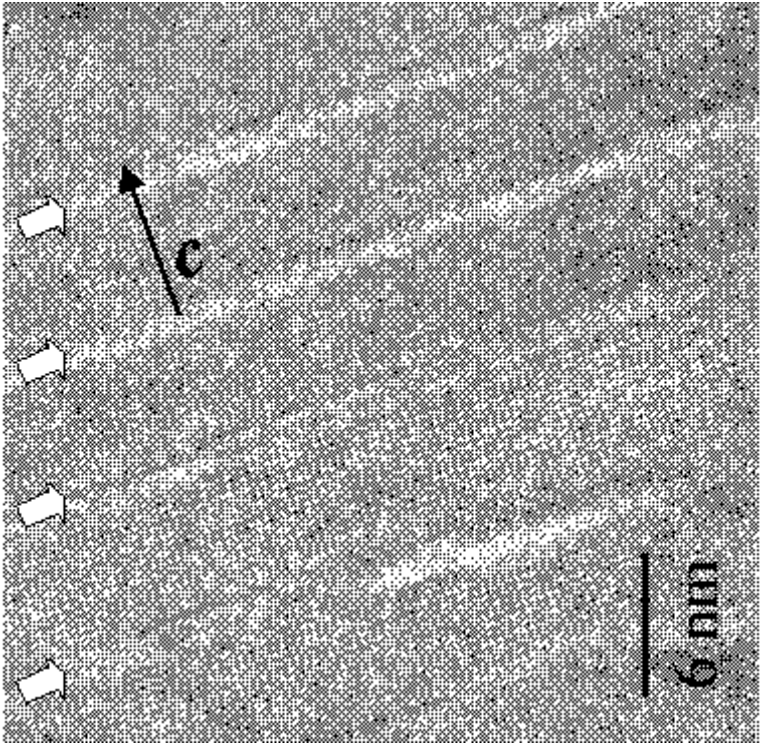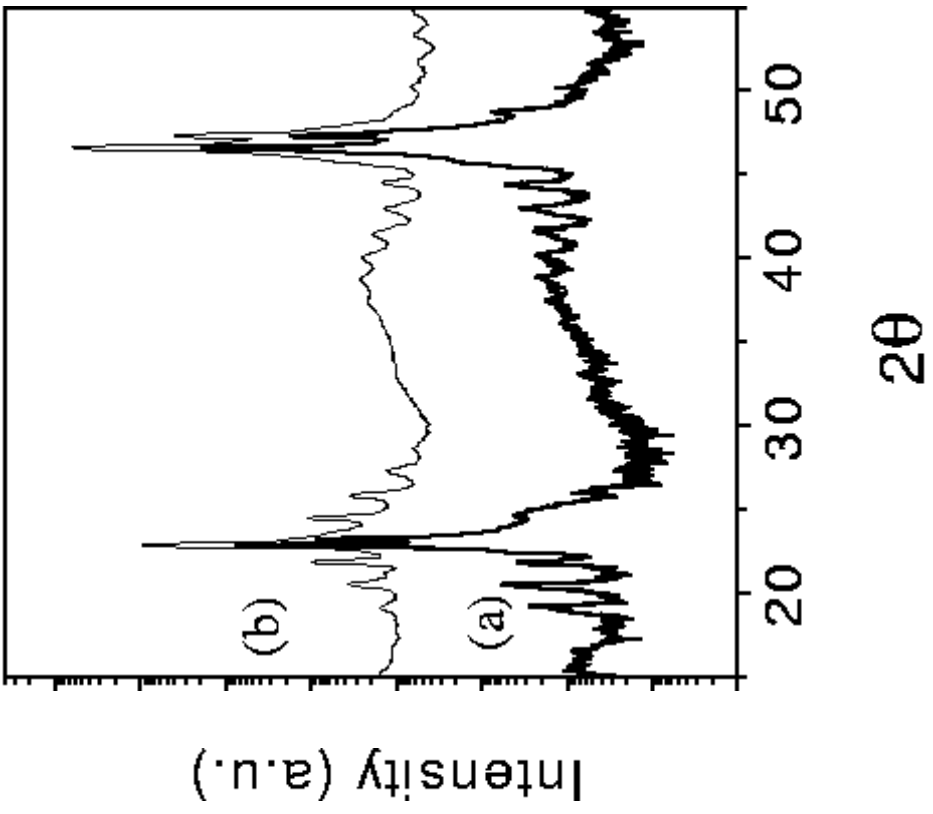

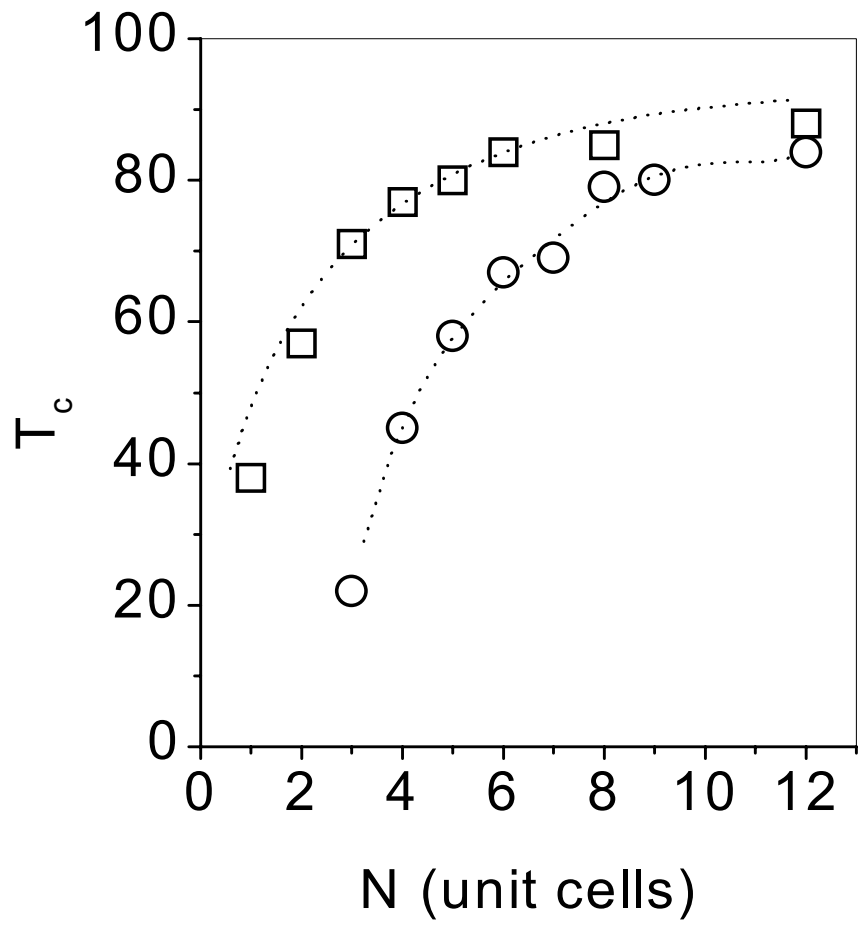

Z. Sefrioui et al figure 5